\documentclass[onecolumn]{aa} 
\usepackage{graphicx}
\usepackage{natbib}
\usepackage{pst-grad}
\usepackage{txfonts}
\newcommand{\diff}{\textrm{d}}
\begin{document}
   \title{A cosmic ray current driven instability in partially ionised
media}


   \author{B. Reville
          \inst{1,2}\fnmsep\thanks{brian.reville@mpi-hd.mpg.de}
          \and
	  J. G. Kirk\inst{1}
          \and
	  P. Duffy\inst{2}
	  \and
	  S. O Sullivan\inst{2}
	  }

   \offprints{B. Reville}

   \institute{Max-Planck-Institut f\"ur Kernphysik, Heidelberg 69029
         \and
         UCD School of Mathematical Science,Belfield, Dublin 4
             }


 
  \abstract
   {We investigate the growth of hydromagnetic waves driven by streaming
   cosmic rays in the precursor environment of a supernova remnant shock.}
   {It is known that transverse waves propagating parallel to the mean magnetic field
    are unstable to anisotropies in the cosmic ray distribution, and may provide
    a mechanism to substantially amplify the ambient magnetic field. We quantify the
    extent to which temperature and ionisation fractions modify this picture.}
   {Using a kinetic description of the plasma we derive the dispersion relation
    for a collisionless thermal plasma with a streaming cosmic ray current.
    Fluid equations are then used to discuss the effects of neutral-ion 
    collisions.}
   {We calculate the extent to which the environment into which the cosmic rays
    propagate influences the growth of the magnetic field, and determines the
    range of possible growth rates.}
   { If the cosmic ray acceleration is efficient, we find that 
    very large neutral fractions are required to
    stabilise the growth of the non-resonant mode. For typical supernova
    parameters in our galaxy, thermal effects do not significantly alter
    the growth rates.
    For weakly driven modes, ion-neutral damping can dominate over the instability
    at more modest ionisation fractions. In the case of a supernova shock 
    interacting with a molecular clouds, such as in RX J1713.7-3946, with high
    density and low ionisation, the modes can be rapidly damped.
    }


   \maketitle
%
%


\section{Introduction}

There is clear evidence from radio, X-ray and gamma ray observations
that electrons, and probably protons, are accelerated in supernova 
remnants (SNR). First order Fermi acceleration is believed to be 
the most likely mechanism for this acceleration,
as it provides a natural explanation for the spectral shape
(for a recent review see~\citealt{Hillas}).
Cosmic rays up to the ankle are believed to have their origin in
our galaxy. However, the standard picture of Fermi
acceleration must be stretched to its limit in order to
accelerate protons to even the knee of
the spectrum at $\sim 3\times 10^{15}~{\rm eV}$ \citep{LagageCesarsky}.
Various methods have been suggested to solve this problem 
in the quasilinear framework, but amplification 
of the ambient magnetic field probably provides the best 
possibility~\citep[e.g.][]{KirkDendy}.
The presence of bright X-ray rims near the shocks of many young
supernova remnants has been interpreted as observational 
evidence of field amplification (\citealt{VinkLaming, voelk, Ballet}), but 
more detailed comparisons of these data with radio observations are
necessary to confirm this result \citep{KatzWaxman}.

Recently an attempt to derive a non-linear theory of 
magnetic field amplification was made by~\cite{BellLucek},
and supported by hybrid numerical simulations \citep{LucekBell}.
It was shown
that the cosmic ray pressure could wind up the frozen-in magnetic
field to values in excess of the typically assumed self-quenching
level of $\delta B/B_0 \lesssim 1$. Building on these papers, \cite{Bell2004}
reanalysed the work of~\cite{acht83},
and found a strong non-resonant mode that had been previously
overlooked. Numerical MHD simulations in \cite{Bell2004, Bell2005}
verified that the streaming
of cosmic rays can amplify the magnetic field at short lengthscales
to values greater than the ambient field. 
The instability is driven at short wavelengths due to the thermal
plasma attempting to neutralise the effects of the streaming cosmic rays.
This induces a current in the thermal plasma. The cosmic rays
essentially behave as unmagnetised particles and do not
move coherently with magnetised thermal background particles.
The same instability has also been seen, using particle in cell simulations,~\citep{Niemiec}.
These authors observe the non resonant instability with a considerably
slower growth rate, but this may be due to the artificially
low proton to electron mass ratio ($m_{\rm p}/m_{\rm e} = 10$) 
used in their simulations. This considerably reduces
the range over which modes can be unstable. More realistic mass
ratios 
are required to understand the true nonlinear development of this
instability.

The upstream medium through which the shock-accelerated particles
stream, will, in general, be incompletely ionised. Limits on the
acceleration of cosmic rays in these environments have been 
calculated by \cite{DruryDuffyKirk}, including the damping 
of resonantly excited Alfv\'en waves
due to friction with the neutral particles. This will
also play a significant role in molecular clouds where the ionisation
fractions are even lower and densities much higher. The propagation 
of Alfv\'en waves in molecular clouds has been studied by
\cite{Tagger}. Extending the work of \cite{KulsrudPearse} and
\cite{ZweibelShull}, they
demonstrated that the ponderomotive force of the waves can 
have a significant effect on the structure and resulting magnetic field 
within such clouds.

Recently, \cite{Bykov}, have investigated a similar instability
to that in \cite{Bell2004}, driven by a 
neutralising return current in 
the thermal background plasma, in a partially ionised
medium. A single fluid MHD description of the background particles 
with a generalised Ohm's law is used, and the cosmic ray current
is found by solving the kinetic transport equation. 
They conclude that a non-resonant mode
with a rapid growth rate remains. However, 
as we show in section 3, for high frequency waves
the friction between the charged species and the neutrals is
not sufficient to couple their motions, and a single fluid
description of the plasma is no longer appropriate.

In the present work, we investigate various conditions that may
reduce the growth of the non-resonant instability. In section 2
we follow the analysis of~\cite{acht83}, neglecting collisions,
and quantify the extent to which thermal effects can reduce the
growth of the instability. In section 3 
we introduce a two fluid description 
of the plasma which includes the collisions between the different
species. The neutrals are coupled to the ions through the ion-neutral
friction, and the cosmic rays are coupled to the MHD fluid through
Maxwell's equations. Section 4 applies our results to SN1006 and 
RX~J1713.7-3946. 
We conclude with a discussion about some of the consequences of the
instability in various conditions.


\section{The kinetic non-resonant current driven instability}

The linear dispersion relation for circularly polarised transverse 
waves propagating parallel to the zeroth order magnetic field is
\begin{equation}
\label{disprel}
\frac{c^2k^2}{\omega^2}-1 = \sum_s \chi_s(k,\omega) ,
\end{equation}
where the summation is over each species with corresponding 
charge $q_s$, 
cyclotron frequency $\omega_{\rm cs} = q_{s}B_0/m_{s}c$
and plasma frequency $\omega^2_{\rm ps} = 4\pi e^2 n_{s}/m_{s}$.
The susceptibility $\chi_s$ for each 
component of a weakly ionised plasma is determined by integrating
the Boltzmann transport equation along the unperturbed trajectories
about the zeroth order field (e.g. \citealt{KrallTrivelpiece}).

Previous work on the effect of streaming cosmic rays using a kinetic  
approach has assumed the background plasma to be 
collisionless \citep{acht83,revilleetal,amatoblasi}. For a three-species plasma
with proton electron background and proton cosmic ray component,
Achterberg determined the background susceptibility for
low frequency waves, $\tilde{\omega}_s \ll \omega_{\rm ci} < |\omega_{ce}|$, 
to be
\begin{eqnarray}
\label{Acht_bg}
\omega^2\chi_{bg} = \frac{c^2}{{\rm v_A}^2}\left[ \tilde{\omega}_{\rm i}^2 -\epsilon
\frac{k^2 V_{\rm ti}^2} {\omega_{\rm ci}}\tilde{\omega}_{\rm i}
-\epsilon \frac{\omega_{\rm ci}kJ_{\rm cr}^1}{n_{\rm i}}\right],
\end{eqnarray}
where $\tilde{\omega}_s = \omega - ku_s$ is the Doppler 
shifted frequency with $u_{\rm s}$ the group drift speed of each species, 
${\rm v_A} = B_0/\sqrt{4\pi n_{\rm i} m_{\rm i}}$ is the Alfv\'en velocity
and $V_{\rm ti}^2 = k_B T_{\rm i}/m_{\rm i}$ the ion thermal velocity. For ease of
notation we have also introduced the parameter $\epsilon$ to describe
the polarisation with $\epsilon=+1$ for right-handed waves and $\epsilon=-1$
for left-handed waves ($\omega>0$).
The requirements of zero net current and quasineutrality in the shock 
precursor, where the proton cosmic ray component is anisotropic,
induces a compensating drift in the background plasma
$J_{\rm cr}^1 = n_{\rm cr}\left(u_{\rm cr}-\omega/k\right)$, where $n_{\rm cr}$ and $u_{\rm cr}$ 
are the cosmic ray number density and bulk speed, respectively.
We note that $n_{\rm cr}$ 
is a function of distance from the shock front, since only high energy cosmic rays are 
expected to penetrate far into the precursor

Using the kinetic treatment of Bell for the cosmic ray current,
the resulting dispersion relation takes the form
\begin{eqnarray}
\label{disp}
\tilde{\omega}_{\rm i}^2
+\epsilon \left(\frac{k^2V^2_{\rm ti}}{\omega_{\rm ci}}\right)
\tilde{\omega}_{\rm i}
-{\rm v_A}^2 k^2-
\epsilon \zeta {\rm v}_{\rm s}^2 \frac{k}{r_{\rm gm}}
\left(\sigma(kr_{\rm gm})-1\right)=0,
\end{eqnarray}
where, 

as appropriate in the rest frame of the upstream plasma in the precursor of a 
shock front, we have identified $u_{\rm cr}$ with the incoming plasma 
speed ${\rm v}_{\rm s}$. 
We define $r_{\rm gm} = p_{\rm min}c/eB_0$ the gyroradius of the 
minimum energy cosmic ray and consider only waves with 
$k>0$. The function $\sigma$ represents
the cosmic ray susceptibility normalised to the
return current $J_{\rm cr}^1$ 
as described in Eq (12) of \cite{Bell2004}. 
We extend this result for arbitrary power law distributions
$f\propto p^{-4+q}$, between $p_{\rm min}$ and $p_{\rm max}$. For $q < 1$ 
\begin{eqnarray}
 \sigma(x) = 
\frac{3}{4}\left(\frac{x^{-3}}{q-4}-\frac{x^{-1}}{q-2}\right)
\ln\left|\frac{x+1}{x-1}\right|-\frac{3x^{-2}}{2(q-4)} +
\;\;\;\;\;\;\;\;\;\;\;\;\;\;\;\;\;\; \nonumber\\
\frac{3}{2(q-1)}\left(\frac{1}{q-4}-\frac{1}{q-2}\right)
\left[\textrm{Re}\left\lbrace{_2F_1}\left(
\frac{1-q}{2},1,\frac{3-q}{2};x^{-2}\right)\right\rbrace -1\right]
\nonumber
\end{eqnarray}
\begin{equation}
\label{sigma}
+\frac{3\pi}{4} i  \epsilon
\left\lbrace \begin{array}{lr}
\frac{x^{-3}}{q-4}-\frac{x^{-1}}{q-2} &  x > 1 \\
\frac{x^{1-q}}{q-4}-\frac{x^{1-q}}{q-2} & x \leq 1
\end{array}
\right.
\end{equation}
where ${_2F_1}(a,b,c;z)$ is the Gauss hypergeometric function.

In (\ref{disp}) we have also 
introduced a dimensionless parameter characterising
the strength of the driving term
\begin{equation}
\zeta = 
\frac{n_{\rm cr} p_{\rm min}}{n_{\rm i} m_{\rm i} {\rm v}_{\rm s}}
\nonumber
\end{equation}

In a shock precursor, we expect $p_{\rm min}$ to increase with distance
ahead of the shock front \citep[e.g.,][]{eichler79,blasi}, since higher 
energy particles have a larger mean free path. For
$q=0$, and neglecting a logarithmic term, this leads to $n_{\rm cr}\propto
1/p_{\rm min}$.  Thus, $\zeta$ is approximately constant, reflecting the
fact that the energy density of the cosmic rays is the same in each
decade of momentum.  It is roughly given by ${\rm v}_{\rm s}/c$ times
the ratio of the cosmic ray energy density at the shock to the
incoming flux of momentum $n_{\rm i} m_{\rm i} {\rm v}_{\rm s}^2$.  If SNR shocks
are efficient in accelerating cosmic rays, one expects 
this ratio to be at least several per cent. \citep[e.g.,][]{Volketal}.
Hence, $\zeta\sim 0.01 {\rm v}_{\rm s}/c$. For a more detailed discussion
of this quantity see section 7 of \citet{Bell2004}.

\begin{figure}
\begin{center}
  \includegraphics[width=0.8\columnwidth]{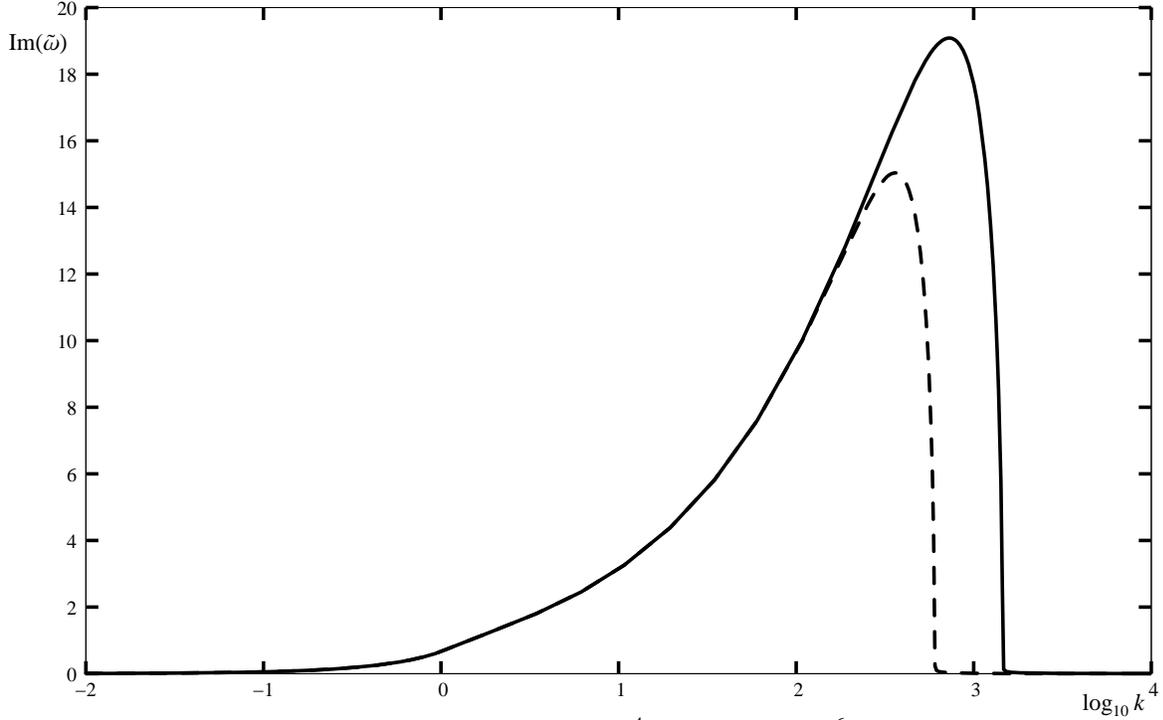}
 \rput(-0.7,-0.1){$\log_{10}k$}
\rput(-15,8.7){Im($\tilde{\omega}$)}
\caption{%
Im($\tilde{\omega}$) for $\epsilon=+1$, plotted for different temperatures.
$T=10^4\,K$ (solid), $T=10^6\,K$ (dashed).
$k$ is plotted in units of 
$r_{\rm gm}^{-1}$ and $\tilde{\omega}_{\rm i}$ in units of ${\rm v}_{\rm s}^2/r_{\rm gm}c$. 
In this plot $r_{\rm gm}=10^{13}{\rm cm}$ and $\zeta = 0.017{\rm v}_{\rm s}/c$
with ${\rm v}_{\rm s}=5000\,$km/s.}
\label{fig_instab}
\end{center}
\end{figure}

The dispersion relation (\ref{disp}) clearly permits waves travelling
in directions parallel and antiparallel to the mean magnetic field for
both left and right polarised waves. For the case of parallel
streaming of the cosmic rays along the magnetic field, the waves that
propagate parallel the field ($\omega > 0$) are unstable, and waves
propagating antiparallel ($\omega < 0$) are damped. This can be
understood physically, as the Lorentz force either acting
centripetally or centrifugally on the field line, depending on the
polarisation of the wave.  

We see from (\ref{sigma}) that $\sigma(kr_{\rm gm})$ is a monotonically decreasing
function of wavenumber for $kr_{\rm gm}>1$. The non-resonant mode emerges when
$\sigma$ is negligible, and the driving term dominates over the \lq\lq
Alfv\'en term\rq\rq\ that contains ${\rm v}_{\rm A}$, 
i.e., when $1 \ll k r_{\rm gm} \ll \zeta {\rm v}_{\rm s}^2/{\rm v}_{\rm A}^2$.

In the frame co-moving with the drifting thermal protons,
the maximum growth rate of the non-resonant mode is
\begin{equation}
\label{gmax}
 {\rm Im}(\omega) = \frac{\zeta}{2}\frac{{\rm v}_{\rm s}}{{\rm v_A}}
\frac{{\rm v}_{\rm s}}{r_{\rm gm}}
\approx\frac{1}{2}\frac{{\rm v}_{\rm s}}{{\rm v_A}}
\frac{n_{\rm cr}}{n_{\rm i}}\omega_{\rm ci},
\end{equation}
which is independent of magnetic field strength.

For a fixed $\zeta$, the growth rate scales with the density as 
$n_{\rm i}^{1/2}$, but the mode grows on shorter length scales as $n_{\rm i}$ increases,
rendering it liable to thermal damping. 
Defining the dimensionless temperature $\Theta = k_BT/m_{\rm i}c^2$,
for large $k$, and again neglecting $\sigma$, it follows from (\ref{disp}),
in the limit $\Theta \gg {\rm v_A}^2/c^2$, that the maximum growth rate is
\begin{equation}
{\rm Im}(\omega) \sim \left({n_{\rm cr}}/{n_{\rm i}}\right)^{2/3}
\left({\beta_s^2}/{\Theta}\right)^{1/3}\omega_{\rm ci},
\end{equation}
where $\beta_s = {\rm v}_{\rm s}/c$. This is the non-relativistic
equivalent of the result found by~\cite{revilleetal}.  At high
temperatures the thermal particles behave as though they were
unmagnetised.  For parameters appropriate for SNR, this effect is shown
in Fig.\ref{fig_instab}. 
In order for the non-resonant mode to leave the 
regime of linear growth before being overtaken by the shock front, i.e., 
before being advected over a distance of roughly $c r_{\rm gm}/{\rm v}_{\rm s}$,
one requires 
${\rm Im}(\omega) > {\rm v}_{\rm s}^2/r_{\rm gm} c$.
The necessary condition for thermal effects to reduce the growth 
rate below this value is
\begin{equation}
 \Theta > \zeta^2 \frac{p_{\rm min}c}{m_{\rm i} {\rm v_s}^{2}}.
\end{equation}
For typical SNR parameters this condition will only be
satisfied for a very weak driving term , $\zeta \ll 1$. 
From Fig.~\ref{fig_instab}
it is clear that damping effects do not suppress the instability 
in the interstellar medium of our
galaxy for typical SNR shock speeds of around $5000\,$km/s.
Thermal effects are likely to play a more significant role
for relativistic shocks, and may even provide a saturation mechanism
for the current driven instability \citep{revilleetal}


\section{The collisional 
non-resonant instability and ion-neutral friction effects}

Unlike the analysis of the previous section, the ideal MHD equations
implicitly include the effects of collisions, although they do not
appear in the final equations. Thus the analysis of \cite{Bell2004}
included the most important effects of Coulomb collisions.  
However, the media into which supernova shocks propagate are generally
not completely ionised, and may even include weakly ionised 
molecular clouds.
Consequently, we now extend the ideal MHD equations to include collisions 
between charged and neutral particles, following the approach of 
\cite{Tagger}, who, in turn, extended the work of \cite{KulsrudPearse}.
Whereas previous work involving ion-neutral
friction has concentrated on shear-Alfv\'en waves, we consider
the non-resonant modes driven by the
streaming cosmic rays. These propagate along the mean field and
are, therefore, circularly polarised. We continue to use
a kinetic description of the cosmic rays,
which remain collisionless. 
Neglecting viscosity and Ohmic friction,
the equations of motion for a frictionally coupled 
system of an MHD fluid and neutrals are given by
\begin{eqnarray}
\label{fluid}
\frac{\diff {\bf u}_{\rm i}}{\diff t} =
\frac{{\bf j \times B}}{\rho_{\rm i} c} - \nu_{\rm in} 
({\bf u}_{\rm i}-{\bf u}_{\rm n}),
\end{eqnarray}
and
\begin{eqnarray}
\label{fluid1}
\frac{\diff {\bf u}_{\rm n}}{\diff t} =
- \nu_{\rm ni} 
({\bf u}_{\rm n}-{\bf u}_{\rm i}),
\end{eqnarray}
\citep{Tagger},
where ${\bf j}$ is the current 
carried by the plasma (excluding the cosmic rays),
${\bf u}_{\rm i}$ and ${\bf u}_{\rm n}$ are the velocities 
of the ionised and neutral components, respectively, and 
$\nu_{\rm in}$ and $\nu_{\rm ni}$ 
are the momentum exchange frequencies.
In writing these equations we have
neglected all terms of order $m_{\rm e}/m_{\rm i,n}$.
The cosmic ray current is contained in Ampere's law
$\nabla \times {\bf B} = (4\pi/c)({\bf j + j}_{\rm cr})$.
The momentum exchange frequencies are related by 
$\rho_{\rm n}\nu_{\rm ni} = \rho_{\rm i}\nu_{\rm in}$, where
$\rho_{\rm n}$ and $\rho_{\rm i}$ are the densities of the neutral and ionised 
components. \cite{KulsrudCesarsky} give the following approximation
for plasmas in the temperature range $10^2~{\rm K} < T < 10^5~{\rm K}$
\begin{equation}
\label{nuin}
 \nu_{\rm in} \approx8.9\times 10^{-9} n_{\rm n}
\left(\frac{T}{10^4~{\rm K}}\right)^{0.4} ~{\rm s}^{-1}.
\end{equation}
The resulting dispersion relation is easily found to be
\begin{equation}
\label{dispIN}
 \omega^2\left(1+\frac{i\nu_{\rm in}}{\omega+i\nu_{\rm ni}}\right)
= k^2{\rm v_A}^2 + \epsilon\zeta\frac{{\rm v}_{\rm s}^2}{r_{\rm gm}}k(\sigma-1).
\end{equation}
In the high frequency limit, $|\omega| \gg \nu_{\rm in}$, this reduces
to a dispersion relation similar to (\ref{disp}), although the
term representing thermal effects is absent. 
In the opposite limit 
$|\omega| \ll \nu_{\rm ni}$ we find
\begin{equation}
\label{denseDR}
 \omega^2 \approx\frac{\rho_{\rm i}}{\rho} 
\left[k^2{\rm v_A}^2 + \epsilon\zeta\frac{{\rm v}_{\rm s}^2}{r_{\rm gm}}k(\sigma-1)\right].
\end{equation}
For high collision frequencies, 
the neutral and ionised components are tied together and the 
effect of ion-neutral collisions is simply 
to increase the effective mass of the ions. 
\begin{figure}
\begin{center}
  \includegraphics[width=0.8\columnwidth]{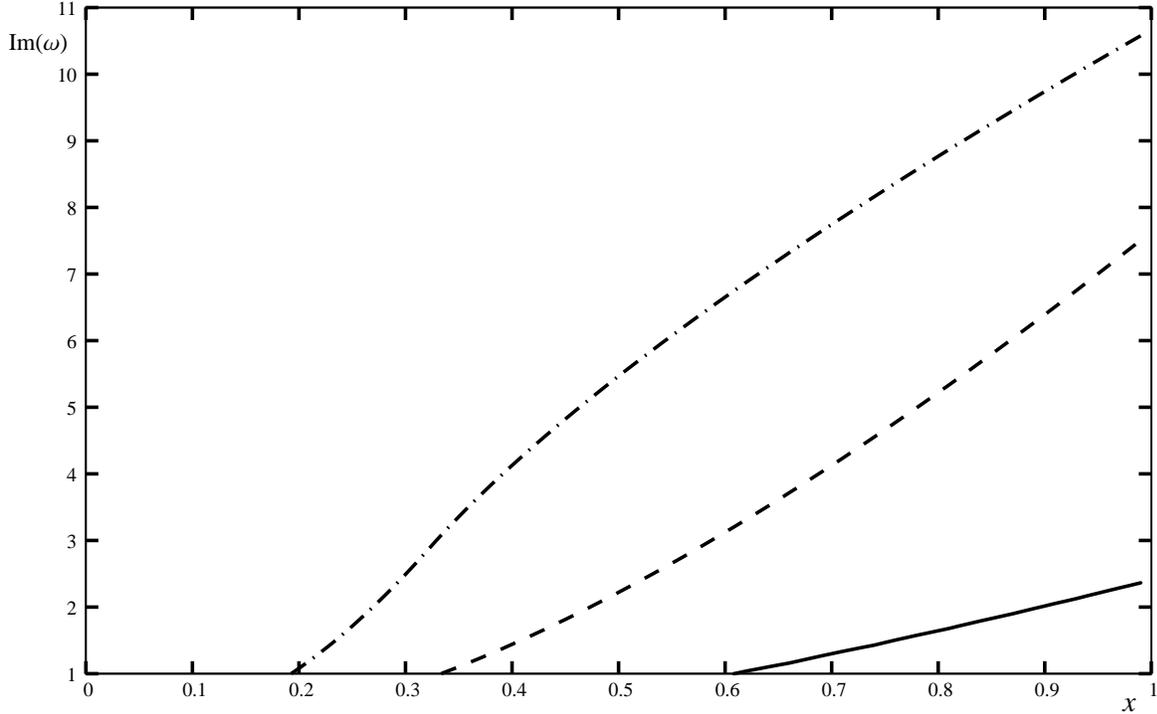}
\rput(-0.5,-0.1){\large{$x$}}
\rput(-15,8.7){Im($\omega$)}
\caption{%
Maximum growth rate as a function
of ionisation fraction for total density $n=n_{\rm i}+n_n$ 
in a H-H$^+$ gas: $0.1{\rm~cm}^{-3}$, $10^4$\,K  (solid),
$1.0{\rm~cm}^{-3}$, $10^3$\,K (dash), $10.0{\rm~cm}^{-3}$, $10^2$\,K (dash-dot).
We take $\zeta=0.01x_{\rm i}{\rm v}_{\rm s}/c$ with
shock speed ${\rm v_s}=5000{\rm ~km~s}^{-1}$, and $r_{\rm gm}=10^{14}{\rm cm}$.
$\omega$ is in units of ${\rm v}_{\rm s}^2/r_{\rm gm}c$. Once the 
$\omega_{\rm max} \leq 1$ the non-resonant instability is not
effective in amplifying the magnetic field. 
}
\label{fig_x}
\end{center}
\end{figure}
 
For strongly driven, non-resonant modes (\ref{dispIN}) reduces to 
\begin{equation}
 \omega(\omega^2+\zeta\frac{{\rm v}_{\rm s}^2}{r_{\rm gm}}k)
+i\nu_{\rm in}\left[(1+Z)\omega^2+Z\zeta\frac{{\rm v}_{\rm s}^2}{r_{\rm gm}}k
\right]=0,
\end{equation}
where $Z=\rho_{\rm i}/\rho_{\rm n} =\nu_{\rm ni}/\nu_{\rm in}$. For
all physically relevant parameters (i.e., $\zeta$, $Z>0$), this cubic
polynomial in $\omega$ has three purely imaginary roots, two of which
are damped modes, and the third the non-resonant growing mode.  Thus,
ion-neutral 
collisions are unable to stabilise the strongly driven mode, although
they can affect its growth rate.

The general analytic expressions for the growing mode are cumbersome, but
various limiting cases yield interesting results.  
In the limit of low ionisation, $Z\gg1$ one recovers
the growth rate given in (\ref{gmax}). 
For $Z\ll1$ the 
growth rate is given by
\begin{equation}
 \gamma=-\frac{\nu_{\rm in}}{2}+
\frac{1}{2}
\sqrt{\nu_{\rm in}^2+4\zeta\frac{{\rm v}_{\rm s}^2}{r_{\rm gm}}k}
\end{equation}
If $\nu_{\rm in}^2\ll4\zeta{\rm v}_{\rm s}^2k/r_{\rm gm}$ collisions are too slow
to compete with the driving, and  
the growth rate is again
the same as (\ref{gmax}) with only a small reduction due to collisions.
However, in addition to the low frequency limit in which the components are
tied together (\ref{denseDR}), a new regime arises. If the collisional
drag on the ions is sufficient to affect the driving, but not to couple the components, i.e., 
$\nu_{\rm in}^2\gg 4\zeta{\rm v}_{\rm s}^2k/r_{\rm gm}\gg \nu^2_{\rm ni}$, 
then we obtain a similar expression
to that of \cite{Bykov}, 
\begin{equation}
 \gamma\approx \zeta\frac{{\rm v_s}^2}{\nu_{\rm in} r_{\rm gm}}k=
 \frac{\omega_{\rm ci}}{\nu_{\rm in}}\frac{n_{\rm cr}}{n_{\rm i}}k{\rm v_s} .
\end{equation}
However, this applies only when $\gamma\ll\nu_{\rm in}$.

In Fig.~\ref{fig_x} we illustrate the influence of 
ion-neutral collisions for SNR parameters. Thermal damping
is negligible, but a weak dependence on temperature enters via
Eq (\ref{nuin}). 
Using (\ref{dispIN}), we plot the maximum growth
rate, as a function of ionisation fraction $x_{\rm i}\equiv n_{\rm i}/n$,
taking the driving term $\zeta=0.01 x_{\rm i} {\rm v_s} /c$
since the cosmic rays couple directly only to the ionised component.
For higher density plasmas, a larger fraction of neutral 
particles is necessary to reduce the growth rate below the threshold
value of ${\rm v}_{\rm s}^2/r_{\rm gm}c$.


\section{Application to SNR}

X-ray observations of several young supernovae display
bright rims immediately downstream of the outer shock.
This has been modelled as the synchrotron cooling of relativistic
electrons, which requires a magnetic field substantially higher than in the surrounding medium
\citep{VinkLaming, Ballet, voelk}. In the case of 
SN~1006 the magnetic field at the shock front is estimated to be of the order $100 \mu{\rm G}$. 
One way in which field amplification may take place is if the non-resonant instability 
discussed above occurs upstream of the shock front. Since this instability depends on 
strong driving by cosmic rays, it cannot bootstrap from an initial state in which they are absent. 
Nevertheless, we can check its consistency, by requiring that, given a substantial cosmic ray 
pressure at the shock,
the growth rate is sufficiently rapid for it to enter the nonlinear regime before 
being overtaken by the shock. 
Thus, 
the physical conditions far upstream ($B$, $n$, $x_{\rm i}$, $T$) together with the shock speed
${\rm v}_{\rm s}$ and cosmic ray intensity $U_{\rm cr}/\rho_{\rm i}{\rm v_s}^2=0.1$,
should combine to yield a growth rate in excess of ${\rm v_s}^2/(r_{\rm gm}c)$.

In the case of SN1006, the shock velocity is approximately ${\rm v_s} \approx2900{\rm~km~s}^{-1}$
the density is $0.05 \lesssim n \lesssim 0.25 {\rm ~ cm}^{-3}$, and the neutral fraction
is small, $x_{\rm i}\approx0.9$ \citep{Raymond}. Taking 
$n_{\rm i}=0.25 {\rm ~ cm}^{-3}$ gives 
$\zeta{\rm v_s}^2/{\rm v_A}^2 \sim 10$.  The low density and shock speed, 
imply that
the non-resonant mode is not very strongly driven,  
but the driving is nevertheless strong compared to the collision frequency.  
Very close to the shock, the growth time of the field is on the order of
years, while further from the shock 
the growth rate decreases quite dramatically, as
the number density of cosmic rays decreases.
For the parameters given above the maximum growth rate is 
$\sim 10 {\rm v_s}^2/r_{\rm gm} c$. The non-resonant mode may still amplify 
the field into the nonlinear regime.

The low neutral fraction in SN1006 means that ion-neutral collisions
are not likely to
dramatically reduce the instability. The interaction
of a supernova blast wave with a molecular cloud, 
such as that observed close to
the north-western rim of RX~J1713.7-3946
would have a much 
more significant effect on the growth. 
Although the exact parameters are uncertain,
molecular clouds are generally clumpy with interclump densities in the range 
$5-25 {\rm~cm}^{-3}$ and ionisation fractions not larger than $10\%$
\citep{Chevalier99}. Adopting the values in \cite{Malkovetal}
$T=10^2{\rm~K},~n=23{\rm~cm}^{-3}, ~x_{\rm i}=0.01, {\rm~v_s}=10^8{\rm cm~s}^{-1}$
the collision frequency is larger than the growth rate.
The non-resonant driving term is unable to dominate
over the Alfv\'en term and growth is driven resonantly by the
cosmic rays themselves, at very long wavelengths $kr_{\rm gm}\ll1$.
At shorter length scales, the waves are rapidly ion-neutral damped Alfv\'en
waves. The shortest growth timescale is on the order of $\sim 10^6$ yrs. 
This value varies with choice of $r_{\rm gm}$, temperature and density,
but for high density and low ionisation fractions, the growth 
timescale is longer than the free expansion phase of a 
supernova remnant, or the lifetime of the wave in the shock precursor,
suggesting that the Fermi acceleration mechanism 
may switch off upon interaction with a molecular cloud.


\section{Discussion}

It has been suggested that the non-resonant cosmic ray current
driven instability may play a significant role in the acceleration
of cosmic rays \citep{Bell2004}. In the case of strongly driven unstable modes,
the mean field is unable to quench the growth of the waves at
$\delta B \lesssim B_0$. The maximum achievable energy
may be pushed beyond the Lagage-Cesarsky limit~\citep{LagageCesarsky}, 
not only by producing nonlinear turbulence to scatter the particles
but also through amplification of the ambient field.

In this paper we have shown that the non-resonant instability 
can have a very rapid growth rate under most young SNR conditions,
in the interstellar medium. Thermal effects are
of little importance for nonrelativistic shocks, and large
neutral fractions are required to prevent growth beyond the 
linear regime.
However,
efficient injection and acceleration are essential for this
mechanism to occur. The largest uncertainty in our models
arises in determining the strength of the driving term $\zeta$, which 
in our models depends only on ionisation fraction and not
explicitly on the ion density. As
a result the instability increases in more dense regions. 
It should be noted however, 
that energy losses close to the injection momentum
can reduce cosmic ray production in over dense regions 
\citep{DruryDuffyKirk}. 

The nonlinear evolution of field amplification
via this instability requires further study. Current MHD simulations
have only been performed with constant cosmic ray currents 
\citep{Bell2004, Bell2005}, using periodic boundary conditions. 
The first results of PIC simulations of this mechanism have
recently been published~\citep{Niemiec}. The dynamical range
over which the non-resonant instability operates is dependent
on the mass differences between magnetised and unmagnetised
particles. Realistic proton electron mass ratios are 
essential in simulating this process. This is most likely the
reason for the relatively low saturation value of 
$\delta B \sim B_0$.
As the 
field becomes highly nonlinear, $\delta B > B_0$, the cosmic ray 
trajectories will be deflected. Future PIC simulations will
give further insight the evolution and saturation 
of the non-resonant instability.

Both the numerical MHD simulations of \cite{Bell2004,Bell2005}
and the PIC simulations of \cite{Niemiec} 
show that in the initial stages of the instability's development
cavities and filaments are created along the 
direction of the mean magnetic field.  
It is clear from equation (\ref{fluid1}) that the resulting 
cavities would be mostly populated by neutral species. 
This is in contrast with the results of \cite{Tagger}, where the
wave damping expels the neutrals from ionised flux tubes, although
the waves they considered were not driven by any mechanism. 
In highly dense, low ionisation regions, where the instability cannot
exceed the damping, the analysis of \cite{Tagger} applies.

The connection between magnetic field amplification and the broadening
of the narrow component H$\alpha$ line has been discussed by~\cite{Ghavamian}.
These authors suggest that an observed broad component, 
of width $\delta v$, such that $\delta v/{\rm v_A}>1$, implies
a turbulent field component $\delta B/B >1$. We have demonstrated that the 
cosmic ray current driven instability is still capable of amplifying fields
in the Balmer dominated shocks. This may provide a possible observational
test of the amplification of the turbulent 
magnetic component in supernova shocks.

\begin{acknowledgements}
      This research was jointly supported by 
	COSMOGRID and the Max-Planck-Institut f\"ur Kernphysik, Heidelberg.
We thank the referee for several insightful suggestions. 
BR would like to thank Prof. A.R. Bell for many helpful discussions.
\end{acknowledgements}

\end{document}